\documentclass[aps,prb,showpacs,groupedaddress,floatfix,nofootinbib]{revtex4}

\usepackage{graphicx}
\usepackage{color}

\begin{document}

\title{Quantum scattering by Wronskians}
\author{Francisco M. Fern\'andez}\email{fernande@quimica.unlp.edu.ar}

\affiliation{INIFTA (UNLP, CCT La Plata--CONICET), Blvd. 113 y 64
S/N, Sucursal 4, Casilla de Correo 16, 1900 La Plata, Argentina}

\begin{abstract}
We show that Wronskians between properly chosen linearly independent
solutions of the Schr\"odinger equation greatly facilitate the study of
quantum scattering in one dimension. They enable one to obtain the necessary
relationships between the coefficients that determine the asymptotic
behavior of the wavefunction. As illustrative examples we calculate the
transmission probability for the penetration of a Gaussian barrier and the
scattering resonances for a Gaussian well.
\end{abstract}


\maketitle

\section{Introduction}

\label{sec:intro}

Barrier penetration or tunnel effect is one of the most striking
predictions of quantum mechanics and most textbooks discuss the
few available exactly solvable one--dimensional models\cite{T75,
M76, CDL77, F99,PC61}. In this journal there has been great
interest in the subject\cite{C65,
D79,WHZ82,C88,W88,LV91,C00,RMNS02,L03,P02,DK10} as well as in
one--dimensional potential scattering in general\cite{E65,D79,WHZ82,S88,SW96}%
. Such interest has been focussed not only on the analytical properties of
potential scattering\cite{E65,SW96,NR96} but also on the numerical
calculation of transmission probabilities and other quantities that describe
barrier penetration and potential scattering\cite
{C88,LV91,KL91,C00,L03,P02,DK10}.

The purpose of this paper is to show that a straightforward
application of Wronskians, which are well known in the study of
ordinary linear differential equations\cite{A69,PC61}, greatly
facilitates the discussion of the analytical properties of
potential scattering. They are also suitable for the numerical
calculation of the transmission probability and any other quantity
of interest in potential scattering. Present discussion is
motivated by an earlier application of Wronskians to the analysis
of one--dimensional models for resonance tunneling
reactions\cite{WC73}.

In Sec.~\ref{sec:Schro} we apply Wronskians to the Schr\"{o}dinger equation
for potential scattering in one dimension and derive matrix equations
connecting the coefficients of the asymptotic forms of the wavefunction left
and right of the scattering center. In Sec.~\ref{sec:tunneling} we
specialize in a general short--range interaction and derive equations for
the transmission probability. As illustrative nontrivial examples we
consider a Gassian barrier and a Gaussian well. By means of an
exactly--solvable problem we discuss the origin of the scattering resonances
in potential wells. In Sec.~\ref{sec:conclusions} we summarize the main
results and draw conclusions. In order to make this paper sufficiently
self--contained we collect some well known mathematical properties of the
Wronskians in an Appendix.

\section{Wronskians and the Schr\"{o}dinger equation}

\label{sec:Schro}

The time--independent Schr\"{o}dinger equation for a particle of mass $m$
that moves in one dimension ($-\infty <X<\infty $) under the effect of a
potential $V(X)$ is
\begin{equation}
-\frac{\hbar ^{2}}{2m}\psi ^{\prime \prime }(X)+V(X)\psi (X)=E\psi (X)
\label{eq:Schrodinger}
\end{equation}
If we define the dimensionless coordinate $x=X/L$, where $L$ is an
appropriate length scale, then we obtain the dimensionless eigenvalue
equation
\begin{eqnarray}
&&-\frac{1}{2}\varphi ^{\prime \prime }(x)+v(x)\varphi (x)=\epsilon \varphi
(x)  \nonumber \\
&&\varphi (x)=\sqrt{L}\psi (Lx),\;v(x)=\frac{mL^{2}}{\hbar ^{2}}%
V(Lx),\;\epsilon =\frac{mL^{2}E}{\hbar ^{2}}  \label{eq:Schro_dim}
\end{eqnarray}
The length unit $L$ that renders both $\epsilon $ and $v(x)$ dimensionless
is arbitrary and we can choose it in such a way that makes the
Schr\"{o}dinger equation simpler. We will see some examples in Sec.~\ref
{sec:tunneling}.

In most cases of physical interest we can write the asymptotic behavior of
the dimensionless wavefunction $\varphi (x)$ as follows
\begin{equation}
\varphi (x)\rightarrow \left\{
\begin{array}{c}
A_{1}C_{1}(x)+B_{1}S_{1}(x)\text{ for }x\rightarrow \infty \\
A_{3}C_{3}(x)+B_{3}S_{3}(x)\text{ for }x\rightarrow -\infty
\end{array}
\right.  \label{eq:phi_1,3}
\end{equation}
and in the intermediate region $-\infty <x<\infty $ we have
\begin{equation}
\varphi (x)=A_{2}C_{2}(x)+B_{2}S_{2}(x)  \label{eq:phi_2}
\end{equation}
The form of the functions $C_{j}(x)$ and $S_{j}(x)$ depends on the problem
and throughout this paper we choose all of them real. On the other hand, the
coefficients $A_{j}$ and $B_{j}$ may be complex. We discuss some examples in
Sec.~\ref{sec:tunneling}; for the time being we assume that those functions
satisfy the conditions (\ref{eq:C,S}) at a conveniently chosen point $x_{0}$%
. By means of the Eq. (\ref{eq:A,B_W}) given in the Appendix we can easily
obtain matrix expressions connecting the coefficients $A_{i}$ and $B_{i}$ in
the three regions:
\begin{eqnarray}
\left(
\begin{array}{c}
A_{1} \\
B_{1}
\end{array}
\right) &=&\mathbf{R}_{1}\left(
\begin{array}{c}
A_{2} \\
B_{2}
\end{array}
\right) ,\;\left(
\begin{array}{c}
A_{2} \\
B_{2}
\end{array}
\right) =\mathbf{R}_{2}\left(
\begin{array}{c}
A_{3} \\
B_{3}
\end{array}
\right) ,  \nonumber \\
\left(
\begin{array}{c}
A_{1} \\
B_{1}
\end{array}
\right) &=&\mathbf{R}\left(
\begin{array}{c}
A_{3} \\
B_{3}
\end{array}
\right) ,\;\mathbf{R=R}_{1}\cdot \mathbf{R}_{2}  \label{eq:connect_AB}
\end{eqnarray}
where
\begin{eqnarray}
\mathbf{R}_{1} &=&\left(
\begin{array}{ll}
W(C_{2},S_{1}) & W(S_{2},S_{1}) \\
W(C_{1},C_{2}) & W(C_{1},S_{2})
\end{array}
\right) ,  \nonumber \\
\mathbf{R}_{2} &=&\left(
\begin{array}{ll}
W(C_{3},S_{2}) & W(S_{3},S_{2}) \\
W(C_{2},C_{3}) & W(C_{2},S_{3})
\end{array}
\right)  \label{eq:R1,R2}
\end{eqnarray}
In these equations $W(f,g)$ denotes the Wronskian of the functions $%
f(x) $ and $g(x)$\cite{PC61,A69} already defined in the Appendix.
If we repeat the procedure and obtain the inverse relations we
appreciate that
\begin{eqnarray}
\mathbf{R}_{1}^{-1} &=&\left(
\begin{array}{ll}
W(C_{1},S_{2}) & -W(S_{2},S_{1}) \\
-W(C_{1},C_{2}) & W(C_{2},S_{1})
\end{array}
\right) ,  \nonumber \\
\mathbf{R}_{2}^{-1} &=&\left(
\begin{array}{ll}
W(C_{2},S_{3}) & -W(S_{3},S_{2}) \\
-W(C_{2},C_{3}) & W(C_{3},S_{2})
\end{array}
\right)  \label{eq:R1,R2_inv}
\end{eqnarray}
from which it follows that
\begin{equation}
\mathbf{R}_{j}^{t}\mathbf{J}=\mathbf{JR}_{j}^{-1},\;\mathbf{J}=\left(
\begin{array}{ll}
0 & 1 \\
-1 & 0
\end{array}
\right)  \label{eq:Rj_trans-Rj_inv}
\end{equation}
where the superscript $t$ stands for transpose. Since the skew--symmetric
matrix $\mathbf{J}$ satisfies $\mathbf{J}^{t}=\mathbf{J}^{-1}=-\mathbf{J}$
we conclude that
\begin{equation}
\mathbf{R}^{t}\mathbf{J}=\mathbf{JR}^{-1}  \label{eq:R_trans-R_inv}
\end{equation}
which resembles the symplectic condition for a canonical transformation in
classical mechanics\cite{G80}. Any matrix that satisfies Eq.~(\ref
{eq:R_trans-R_inv}) is said to be symplectic\cite{G80}. Besides, Eq. (\ref
{eq:R1,R2_inv}) tells us that the determinant of every symplectic matrix $%
\mathbf{R}_{1}$, $\mathbf{R}_{2}$ and $\mathbf{R}$ is unity.

If the potential is parity--invariant ($v(-x)=v(x)$) then $%
C_{2}(-x)=C_{2}(x) $, $S_{2}(-x)=-S_{2}(x)$ and
\begin{eqnarray}
W(C_{2},S_{1}) &=&W(C_{2},S_{3}),  \nonumber \\
W(S_{2},S_{1}) &=&W(S_{3},S_{2})  \nonumber \\
W(C_{1},C_{2}) &=&W(C_{2},C_{3})  \nonumber \\
W(C_{1},S_{2}) &=&W(C_{3},S_{2})  \label{eq:W_sym}
\end{eqnarray}
so that we need to calculate half the number of Wronskians. It is
worth noting that Wronskians containing $C_1$ and $S_1$ are
constant for $x\rightarrow\infty$ and those with $C_3$ and $S_3$
are constant for $x\rightarrow -\infty$.

\section{Potential scattering}

\label{sec:tunneling}

We assume that
\begin{eqnarray}
\lim_{x\rightarrow -\infty }v(x) &=&v_{\_}  \nonumber \\
\lim_{x\rightarrow \infty }v(x) &=&v_{+}  \label{eq:v_asymp}
\end{eqnarray}
where $v_{\pm }$ are finite constants. If $v(x)$ approaches those limits
sufficiently fast then we know that the asymptotic behavior of the solution
is
\begin{eqnarray}
\varphi (x) &\rightarrow &\left\{
\begin{array}{c}
A_{1}^{\prime }e^{ik_{1}x}+B_{1}^{\prime }e^{-ik_{1}x}\text{ for }%
x\rightarrow -\infty \\
A_{3}^{\prime }e^{ik_{1}x}+B_{3}^{\prime }e^{-ik_{1}x}\text{ for }%
x\rightarrow \infty
\end{array}
\right. ,  \nonumber \\
\;k_{1} &=&\sqrt{2(\epsilon -v_{-})},\;k_{3}=\sqrt{2(\epsilon -v_{+})}
\label{eq:phi_1,3_exp}
\end{eqnarray}
provided that $\epsilon >\max \{v_{+},v_{-}\}$.

If we choose
\begin{equation}
C_{j}=\cos (k_{j}x),\;S_{j}(x)=\frac{\sin (k_{j}x)}{k_{j}},\;j=1,3
\label{eq:C_i,S_i}
\end{equation}
in Eq.~(\ref{eq:phi_1,3}) and compare it with Eq.~(\ref{eq:phi_1,3_exp}) we
obtain
\begin{equation}
A_{j}^{\prime }=\frac{1}{2}\left( A_{j}-i\frac{B_{j}}{k_{j}}\right)
,\;B_{j}^{\prime }=\frac{1}{2}\left( A_{j}+i\frac{B_{j}}{k_{j}}\right)
\label{eq:A'B'}
\end{equation}

Suppose that we want to study the scattering of a particle that comes from
the left ($x<0$). In such a case $B_{3}^{\prime }=0$ and the transmission
probability is given by
\begin{equation}
T=\frac{k_{3}|A_{3}^{\prime }|^{2}}{k_{1}|A_{1}^{\prime }|^{2}}  \label{eq:T}
\end{equation}
Noting that $B_{3}=ik_{3}A_{3}$ we can easily rewrite the transmission
probability in terms of the coefficients $A_{j}$ and $B_{j}$ because
\begin{equation}
\frac{A_{3}^{\prime }}{A_{1}^{\prime }}=\frac{2k_{1}A_{3}}{k_{1}A_{1}-iB_{1}}
\label{eq:A'3/A'1}
\end{equation}

In the intermediate or scattering region we write $\varphi (x)$ as in Eq.~(%
\ref{eq:phi_2}) where $C_{2}(x)$ and $S_{2}(x)$ are two solutions of the
dimensionless Schr\"{o}dinger equation that satisfy Eq.~(\ref{eq:C,S}). In
order to obtain them we may resort to any available numerical integration
method, like, for example, Runge--Kutta\cite{PFTV86} (see also
http://en.wikipedia.org/wiki/Runge--Kutta\_methods). Let $y(x)$ be either $%
C_{2}(x)$ or $S_{2}(x)$. Typical numerical integration methods yield $y(x)$
at a set of coordinate points $x_{0}\pm jh$ where $j=0,1,\ldots $ and $h$ is
the step size. They simultaneously provide the derivative of the function $%
y^{\prime }(x)$ at the same set of points so that the numerical
calculation of the Wronskians between the intermediate solutions
and the asymptotic ones is straightforward. Numerical integration
methods like Runge--Kutta are available in many commercial and
free softwares so that it is unnecessary to write a computer
program for that purpose. In the present case we resorted to the
fourth--order Runge--Kutta method built in the computer algebra
system Derive (http://www.chartwellyorke.com/derive.html). The
starting point of the integration process requires $y(x_{0})$ and
$y^{\prime }(x_{0})$ that are already known for the functions
$C_{2}(x)$ and $S_{2}(x)$. We propagate the solution left and
right till the Wronskians appearing in the matrices
(\ref{eq:R1,R2}) are constant within a given error. Then we
calculate the coefficients $A_{1}$ and $B_{1}$ in terms of $A_3$
by means of Eq. (\ref{eq:connect_AB}) and finally the transmission
coefficient from equations (\ref{eq:T}) and (\ref{eq:A'3/A'1}).
Note that the application of the numerical integration method is
straightforward because all the functions $C_{j}(x)$ and
$S_{j}(x)$ are real.

In the simplest case of a parity--invariant potential we need half the
Wronskians in order to obtain the matrix $\mathbf{R}$ in equations (\ref
{eq:connect_AB}) and (\ref{eq:R1,R2}) as discussed at the end of Sec.~\ref
{sec:Schro}. The transmission probability reads
\begin{equation}
T=\frac{k_{1}^{2}}{\left[ \left(
W(C_{2},C_{3})^{2}+k_{1}^{2}W(C_2,S_3)^{2}\right) \left(
W(C_{3},S_{2})^{2}+k_{1}^{2}W(S_3,S_{2})^{2}\right) \right] }
\label{eq:T_sym}
\end{equation}
and we only have to integrate the differential equations for $C_{2}(x)$ and $%
S_{2}(x)$  from $x_{0}=0$ to the right: $x_{j}=jh$, $j=0,1,\ldots ,N$.

As an illustrative example we choose the Gaussian barrier
\begin{equation}
V(X)=V_{0}e^{-\alpha X^{2}},\;V_{0},\alpha >0  \label{eq:V(X)_Gaussian}
\end{equation}
If we set $L=1/\sqrt{\alpha }$ then $v(x)=v_{0}e^{-x^{2}}$ and the
dimensionless Schr\"{o}dinger equation depends on just one potential
parameter $v_{0}=mV_{0}/(\hbar ^{2}\alpha )$. Since $v_{-}=v_{+}=0$ then $%
k_{1}^{2}=2\epsilon $, where $\epsilon =mE/(\hbar ^{2}\alpha )$ is the
dimensionless energy.

\begin{figure}[H]
\begin{center}
\par
\includegraphics[width=9cm]{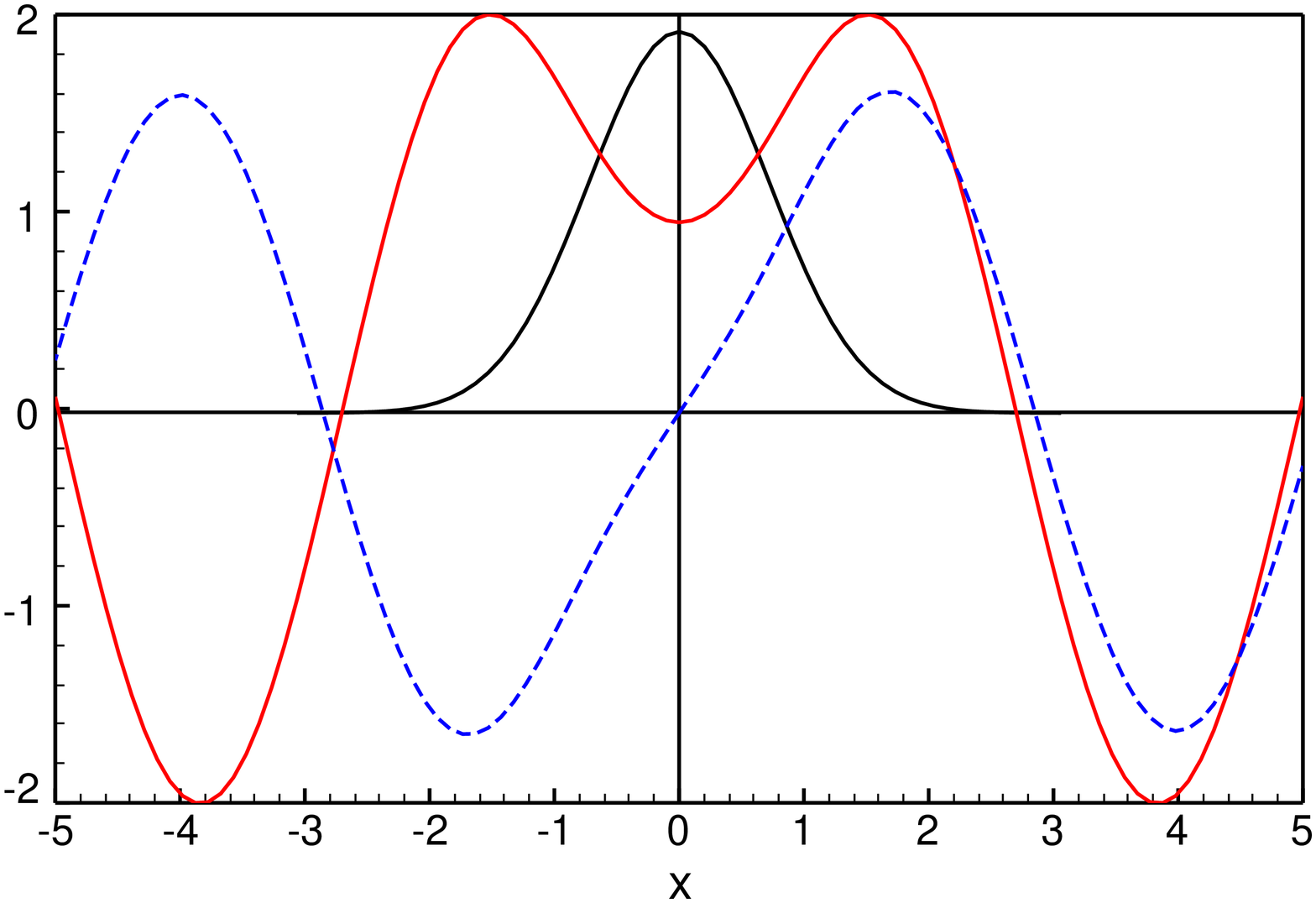}
\par

\par
\end{center}
\caption{Functions $C_2(x)$ (solid line) and $S_2(x)$ (dashed line) for $%
\epsilon=1$ and the Gaussian barrier with $v_0=2$ (solid line)}
\label{fig:C2S2}
\end{figure}

\begin{figure}[H]
\begin{center}
\par
\includegraphics[width=9cm]{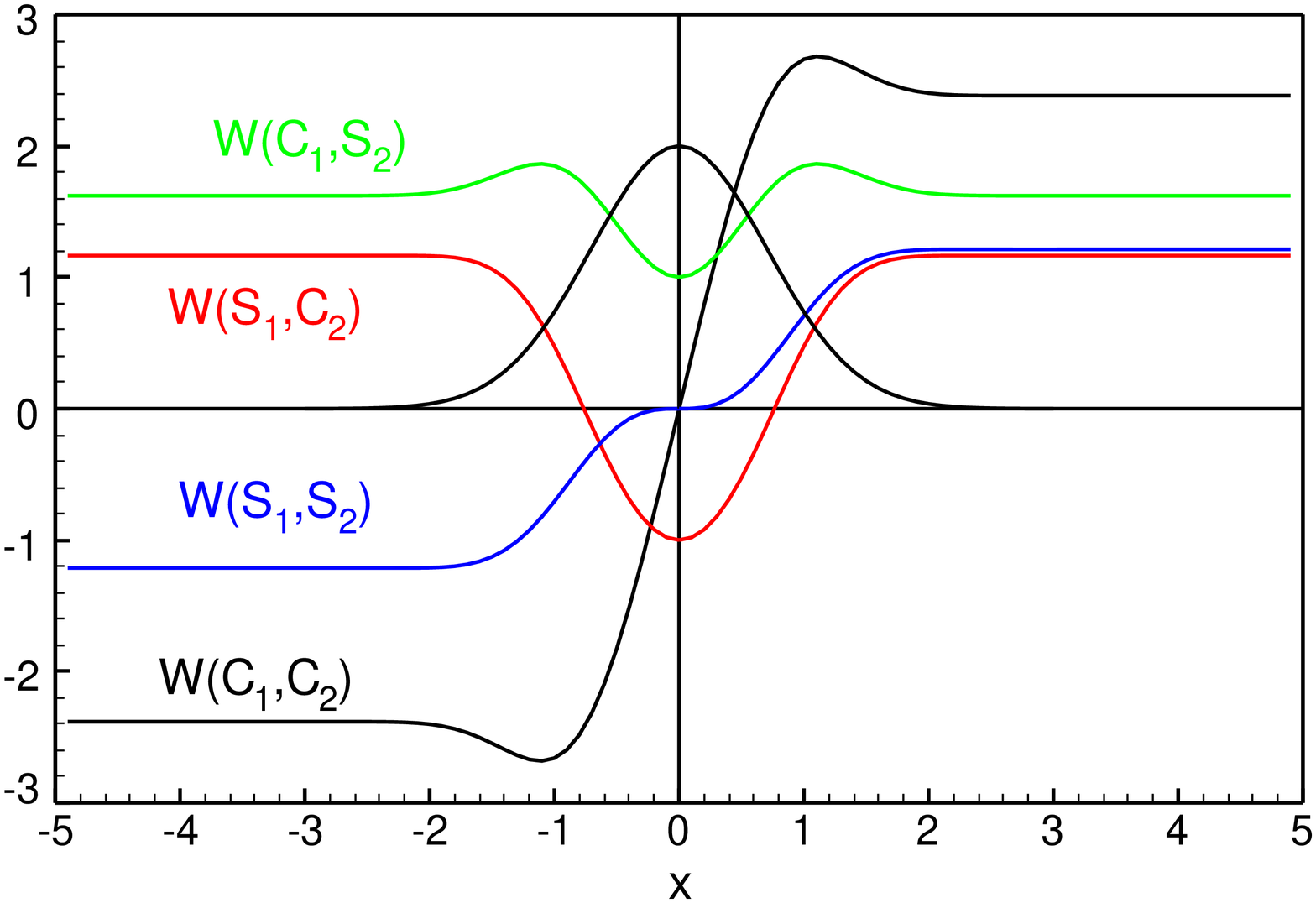}
\par

\par
\end{center}
\caption{Wronskians for $\epsilon=1$ and the Gaussian barrier with
$v_0=2$ } \label{fig:Wronskians}
\end{figure}

\begin{figure}[H]
\begin{center}
\par
\includegraphics[width=9cm]{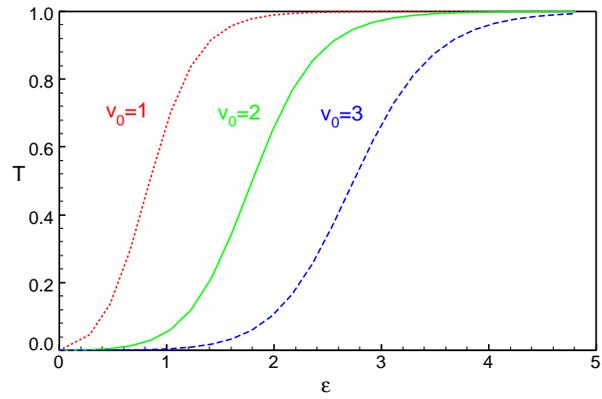}
\par

\par
\end{center}
\caption{Transmission probability for three Gaussian barriers}
\label{fig:TWG}
\end{figure}

Throughout this paper we choose the integration step size $h=0.01$ and $N=500
$ integration points so that the maximum coodinate value is $x_{500}=5$.
Fig.~\ref{fig:C2S2} shows the behaviour of the functions $C_{2}(x)$ and $%
S_{2}(x)$ for $\epsilon =1$ and the Gaussian barrier with $v_{0}=2$. In Fig.~%
\ref{fig:Wronskians} we appreciate that the Wronskians approach
constants as $|x|\rightarrow \infty $ and that $x=5$ is large
enough for an accurate estimation of those limits. We thus
calculated the transmission coefficients shown in Fig.~
\ref{fig:TWG} for three values of $v_{0}$. As expected there is
tunneling for all $\epsilon >0$ and $\lim_{\epsilon \rightarrow
\infty }T=1$. It is clear that the oscillatory behavior of
$T(\epsilon )$ found by Chalk\cite {C88} is due to the truncation
of the Gaussian potential with the purpose of connecting the
power--series solution for the intermediate region with the
asymptotic plane waves\cite{W88}. We appreciate that the necessary
truncation of the integration interval does not produce any
undesirable effect on the transmission probabilities calculated in
terms of Wronskians. In fact, the great advantage of the Wronskian
method is that we calculate the constant asymptotic limit of each
Wronskian with a given desired accuracy as shown in
Fig.~\ref{fig:Wronskians}.

In passing we point out that there is no trace of the questionable tunneling
condition derived by Nandi\cite{N10} for the Gaussia barrier.

\begin{figure}[H]
\begin{center}
\par
\includegraphics[width=9cm]{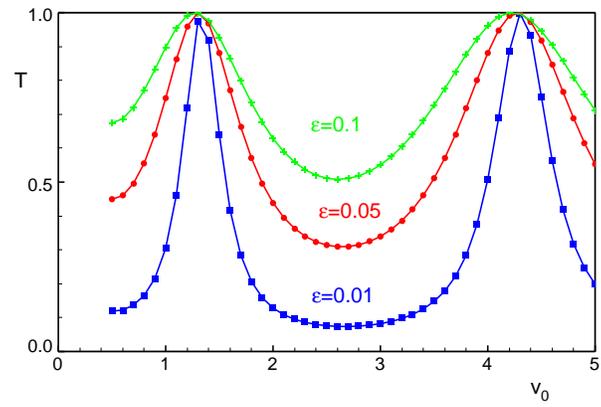}
\par

\par
\end{center}
\caption{Transmission probability for the Gaussian well as a
function of the well depth for three values of the energy}
\label{fig:TWGW}
\end{figure}

Another simple, nontrivial, and most interesting problem is the Gaussian
well
\begin{equation}
V(X)=-V_{0}e^{-\alpha X^{2}},\;V_{0},\alpha >0  \label{eq:V(X)_GW}
\end{equation}
Proceeding as in the preceding example we obtain the dimensionless potential
$v(x)=-v_{0}e^{-x^{2}}$ and the same expressions for $v_{0}$ and $\epsilon $%
. In this case it is most instructive to calculate the transmission
probability $T$ in terms of $v_{0}$ for fixed values of $\epsilon $ in order
to reveal the scattering resonances as shown in Fig.~\ref{fig:TWGW}. Note
that the maxima of the transmission probability $T=1$ that occur at some
particular values of $v_{0}$ are roughly independent of the energy $\epsilon
$. This well known phenomenon is better understood by means of an exactly
solvable model.

The scattering resonances appearing in Fig.~\ref{fig:TWGW} are similar to
the ones exhibited by the exactly solvable well
\begin{equation}
V(X)=-\frac{V_{0}}{\cosh (\alpha X)^{2}},\;V_{0},\alpha >0
\end{equation}
that we easily transform into the dimensionless potential $v(x)=-v_{0}/\cosh
(x)^{2}$. Both the dimensionless potential parameter $v_{0}$ and energy $%
\epsilon $ have the same expressions as in the preceding examples. In this
case the transmission probability reads\cite{F99}
\begin{equation}
T=\frac{\sinh \left( \pi \sqrt{2\epsilon }\right) ^{2}}{\sinh \left( \pi
\sqrt{2\epsilon }\right) ^{2}+\sin (\pi \lambda )^{2}},\;\lambda =\frac{1}{2}%
\left( 1+\sqrt{1+8v_{0}}\right)
\end{equation}
The family of curves $T$ vs. $v_{0}$ for constant $\epsilon $ resembles the
one shown in Fig.~\ref{fig:TWGW}. Note that there is full transmission $T=1$
when $\sin (\pi \lambda )=0$; that is to say, when $\lambda $ is an integer.
In order to understand the origin of these resonances we pay attention to
the bound states\cite{F99}
\begin{equation}
\epsilon _{n}=-\frac{1}{2}(\lambda -1-n)^{2},\;n=0,1,\ldots \leq \lambda -1
\end{equation}
We appreciate that full transmission takes place when one of the
excited bound--state energies ($n>0$) lies exactly at the rim of
the well $\epsilon _{n}=0$ in which case $\lambda =n+1$ (the
ground--state energy $\epsilon_0=-(\lambda-1)^2/2$ is negative for
all $v_0>0$).

This exactly solvable problem also proved to be useful for testing the
accuracy of our programs for the calculation of the transmission probability.

\section{Conclusions}

\label{sec:conclusions}

In this paper we propose an alternative way of approaching quantum
scattering in one dimension. We think that the method based on the
Wronskians between linearly independent solutions to the Schr\"{o}dinger
equation is preferable to other approaches. The relatively light effort
necessary to master a few mathematical properties of the Wronskians pays
generously when attacking the scattering problem either analytically or
numerically. In this paper we focused mainly on the latter because we are
interested in nontrivial problems that are not so widely discussed in most
textbooks on quantum mechanics\cite{T75, M76, CDL77, F99}.

The derivation of all the necessary scattering equations in terms of
Wronskians is straightforward as well as their practical application by
means of extremely simple computer programs. The calculation of the
transmission probability is quite reliable if one simply checks for constant
Wronskians before truncating the propagation of the solutions towards left
and right in the numerical integration routine. Thus, the error due to a
finite integration interval is simply the error in the lack of constant
Wronskians that is easily bounded to the desired accuracy. In this way we
avoid any spurious oscillations in the transmission probability as discussed
in Sec.~\ref{sec:tunneling}.

In addition to all that, the Wronskians have proved to be most
useful for the estimation of the complex energies that describe
tunnel resonances as discussed in the paper on which we based
present pedagogical presentation\cite{WC73}. One can also apply
the Wronskian method to bound states and calculate their energies
by simply taking into account the appropriate asymptotic behavior
of the wavefunction and requiring that it be square integrable. It
is clear that the approach exhibits a wide variety of useful
applications and for that reason we think that it is worth
teaching in advanced undergraduate or graduate courses on quantum
mechanics.

\section{Appendix}

In order to make this paper sufficiently self--contained in this
appendix we outline some well known results about the Wronskians
that are useful for the study of ordinary differential equations
in general\cite{A69} and also for the treatment of the
Schr\"{o}dinger equation in particular\cite{PC61,WC73}. To this
end, we consider the ordinary second--order differential equation
\begin{equation}
L(y)=y^{\prime \prime }(x)+Q(x)y(x)=0  \label{eq:diffeq}
\end{equation}
If $y_{1}$ and $y_{2}$ are two linearly independent solutions to this
equation then we have
\begin{equation}
y_{1}L(y_{2})-y_{2}L(y_{1})=\frac{d}{dx}W(y_{1},y_{2})=0
\end{equation}
where
\begin{equation}
W(y_{1},y_{2})=y_{1}y_{2}^{\prime }-y_{2}y_{1}^{\prime }
\label{eq:Wronskian}
\end{equation}
is the Wronskian (or Wronskian determinant\cite{A69}). Two obvious
properties are:
\begin{equation}
W(f,g)=-W(g,f),\;W(f,f)=0  \label{eq:Wronsk_prop}
\end{equation}

By linear combination of $y_{1}(x)$ and $y_{2}(x)$ we easily
obtain two new solutions $C(x)$ and $S(x)$ satisfying
\begin{equation}
C(x_{0})=S^{\prime }(x_{0})=1,\;C^{\prime }(x_{0})=S(x_{0})=0\;
\label{eq:C,S}
\end{equation}
at a given point $x_{0}$ so that $W(C,S)=1$ for all $x$. If we write the
general solution to Eq.~(\ref{eq:diffeq}) as
\begin{equation}
y(x)=AC(x)+BS(x)
\end{equation}
then
\begin{equation}
A=W(y,S),\;B=W(C,y)  \label{eq:A,B_W}
\end{equation}
This equation is quite useful for deriving relationships between
the coefficients of the asymptotic expansions of the wavefunction
in different regions of space as shown in sections \ref{sec:Schro}
and \ref{sec:tunneling}. A more detailed discussion of the
Wronskians is available in Powell and Crasemann's book on quantum
mechanics\cite{PC61}.

\end{document}